\documentclass[aps, pra,  10pt, twocolumn, longbibliography, floatfix, titlepage=false]{revtex4-1}

\usepackage[nointlimits]{amsmath}
\usepackage{amssymb, cancel}
\usepackage{graphicx, nicefrac}
\usepackage{subfigure}
\usepackage{color}
\usepackage{txfonts}
\usepackage{mathtools}
\usepackage{hyperref}
\hypersetup{colorlinks=true, citecolor=blue, linkcolor=blue}
\graphicspath{{./figures/}}

\begin{document}
\title{Scalable randomized benchmarking of quantum computers using mirror circuits}
\author{Timothy Proctor}
\author{Stefan Seritan}
\author{Kenneth Rudinger}
\author{Erik Nielsen}
\author{Robin Blume-Kohout}
\author{Kevin Young}
\affiliation{Quantum Performance Laboratory, Sandia National Laboratories, Albuquerque, NM 87185, USA and Livermore, CA 94550, USA}
\date{\today}
\begin{abstract} 
The performance of quantum gates is often assessed using some form of randomized benchmarking. However, the existing methods become infeasible for more than approximately five qubits. Here we show how to use a simple and customizable class of circuits---randomized mirror circuits---to perform scalable, robust, and flexible randomized benchmarking of Clifford gates. We show that this technique approximately estimates the infidelity of an average many-qubit logic layer, and we use simulations of up to 225 qubits with physically realistic error rates in the range 0.1-1\% to demonstrate its scalability. We then use up to 16 physical qubits of a cloud quantum computing platform to demonstrate that our technique can reveal and quantify crosstalk errors in many-qubit circuits.
\end{abstract}
\maketitle

Quantum information processors suffer from a wide variety of errors that must be quantified if their performance is to be understood and improved. A processor's errors are commonly probed using randomized benchmarks that involve running random circuits \cite{emerson2005scalable, emerson2007symmetrized, magesan2011scalable, magesan2012characterizing, knill2008randomized, carignan2015characterizing, cross2016scalable, brown2018randomized, hashagen2018real, magesan2011scalable, magesan2012characterizing, magesan2012efficient, carignan2015characterizing, cross2016scalable, brown2018randomized, hashagen2018real, helsen2018new, Helsen2020-it, Claes2020-cy, McKay2020-no, Helsen2020-mb, Morvan2020-ck, proctor2018direct, boixo2018characterizing, arute2019quantum, Liu2021-ij,  cross2018validating}---e.g., standard randomized benchmarking (RB) \cite{magesan2011scalable, magesan2012characterizing} or one of its many variants \cite{knill2008randomized, carignan2015characterizing, cross2016scalable, brown2018randomized, hashagen2018real, magesan2011scalable, magesan2012characterizing, magesan2012efficient, carignan2015characterizing, cross2016scalable, brown2018randomized, hashagen2018real, helsen2018new, Helsen2020-it, Claes2020-cy, McKay2020-no, Helsen2020-mb, Morvan2020-ck, proctor2018direct} cross-entropy benchmarking \cite{boixo2018characterizing}, or the quantum volume benchmark \cite{cross2018validating}. Randomized benchmarks are appealing because they aggregate many kinds of error into one number that quantifies average performance over a large circuit ensemble. Unlike tomographic techniques \cite{Nielsen2020-lt} that estimate a set of parameters that may be exponentially large in the number of qubits ($n$), randomized benchmarks hold the potential for scalable performance assessment.

Yet current randomized benchmarks have one of two scaling problems. Quantum volume and cross-entropy benchmarking require classical computations that are exponentially expensive in $n$, becoming infeasible beyond $n\sim50$ \cite{boixo2018characterizing, arute2019quantum, Liu2021-ij, cross2018validating}. In contrast, standard RB requires only efficient classical computations but it benchmarks composite gates from the $n$-qubit Clifford group. They require $O(\nicefrac{n^2}{\log n})$ two-qubit gates to implement \cite{aaronson2004improved, patel2003efficient, bravyi2020hadamard}, so the fidelity of a typical $n$-qubit Clifford decreases quickly with $n$. Lower compilation overheads [e.g., $O(\log n)$] are possible with access to many-qubit gates \cite{Grzesiak2021-bu}, but in all realistic architectures the circuit depth required to implement a typical Clifford will increase with the number of qubits. As a result, standard RB has only been implemented on up to three qubits \cite{mckay2017three}, and even its streamlined variant ``direct RB'' (DRB) has only been implemented on up to 5 qubits \cite{proctor2018direct}.

\begin{figure}[t!]
\includegraphics[width=8.5cm]{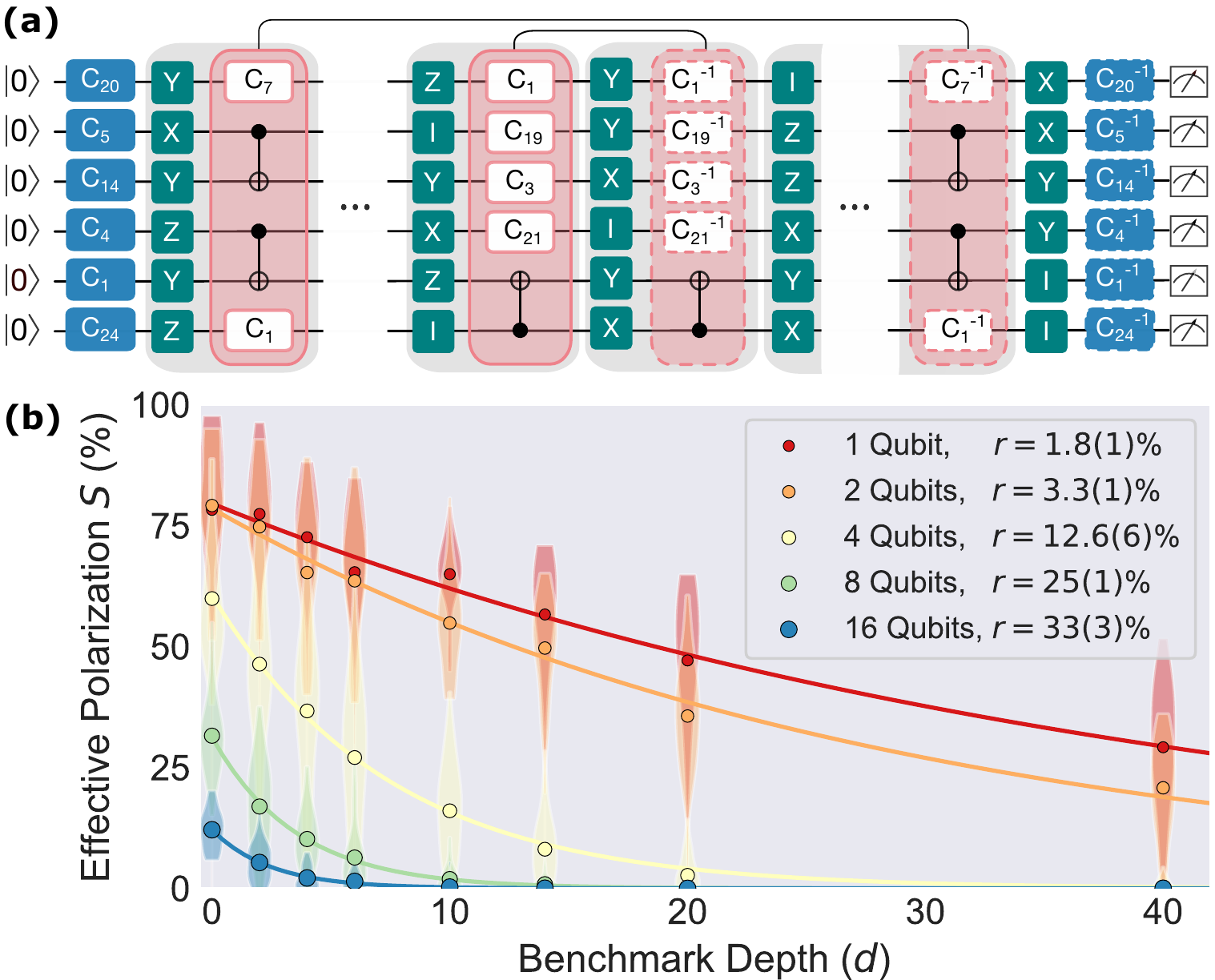} 
\caption{\textbf{Scalable RB with mirror circuits.} \textbf{(a)} Randomized mirror circuits over Clifford gates enable scalable RB. These circuits contain $\nicefrac{d}{2}$ pairs of layers consisting of a layer and its inverse (pink boxes) sampled from some set of $n$-qubit Clifford layers, $d+1$ layers of uniformly random Pauli gates (green boxes), and a layer of uniformly random one-qubit Clifford gates and this layer's inverse (blue boxes). The number of ``Pauli-dressed'' layers (grey boxes) $d$ is the circuit's \emph{benchmark depth}. These circuits' ``effective polarization'' $S$, a quantity closely related to success probability, decays exponential with $d$. \textbf{(b)} Demonstrating our method on 1, 2, 4, 8 and 16 qubit subsets of IBM Q Rueschlikon. Points (violin plots) are the means (distributions) of $S$ versus $d$, and the curves are fits to $S = Ap^d$. Each $r$ is a rescaling of $p$ that approximates the infidelity of an average Pauli-dressed $n$-qubit layer (uncertainties are $1\sigma$ here and thoughout).}
\label{fig:1}
\end{figure}

\begin{figure}[t!]
\includegraphics[width=8.5cm]{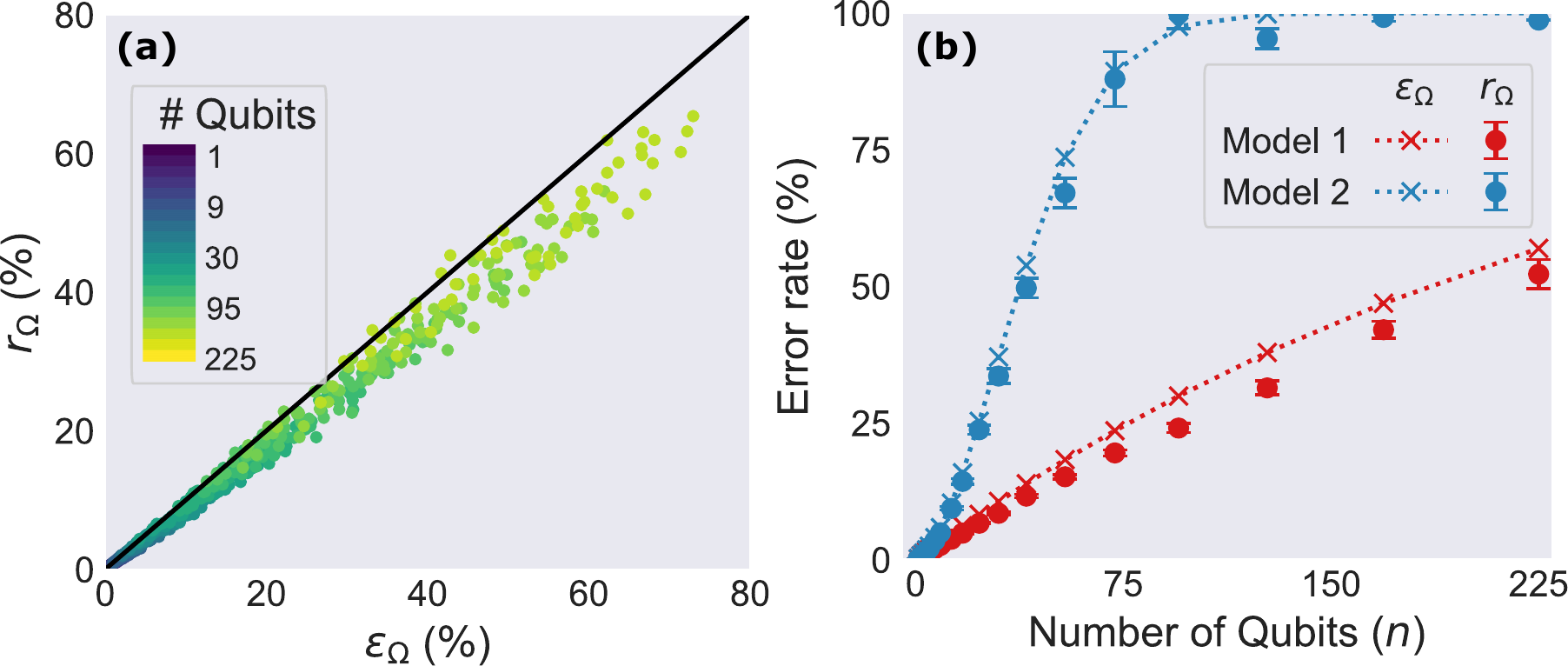}
\caption{\textbf{Validating MRB with many-qubit simulations}. Simulations of MRB on up to 225 qubits show that it reliably approximates the infidelity of $n$-qubit layers, i.e., $r_{\Omega} \approx \epsilon_{\Omega}$.~\textbf{(a)} $r_{\Omega}$ versus $\epsilon_{\Omega}$ for randomly sampled error models. Each point was generated from an independent simulation (sampling an error model and circuits, simulating the circuits, and then applying the analysis to estimate $r_{\Omega}$) for gates subject to stochastic Pauli errors.~\textbf{(b)} $r_{\Omega}$ and $\epsilon_{\Omega}$ versus $n$ for two illustrative error models, with (model 2) and without (model 1) long-range crosstalk. This demonstrates the power of MRB to highlight crosstalk errors.}
\label{fig:sims}
\end{figure}

\begin{figure}[t!]
\includegraphics[width=8.5cm]{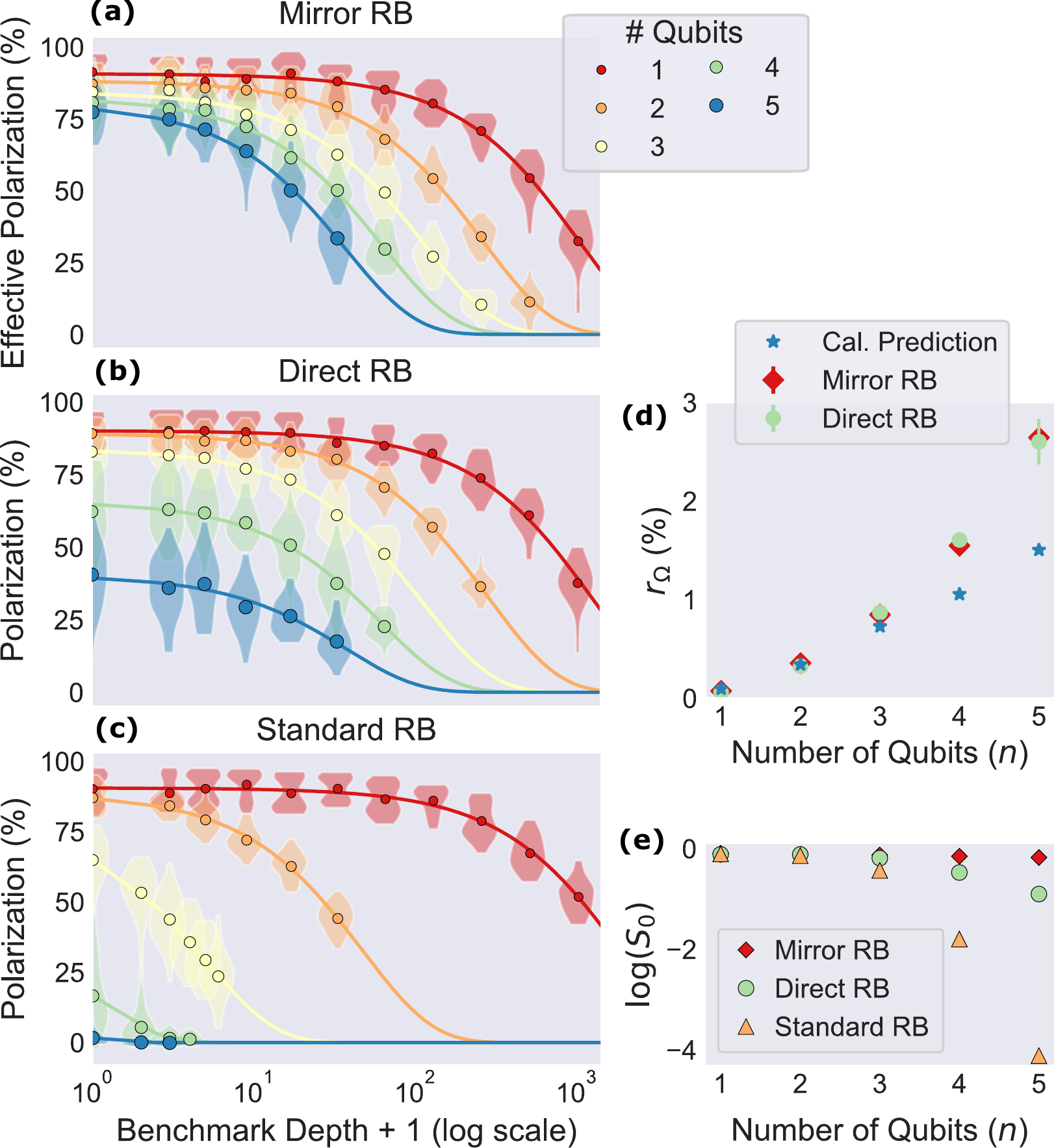}
\caption{\textbf{Validating MRB using cloud access experiments.} MRB, DRB and standard RB on 1-5 qubits of IBM Q Quito. \textbf{(a-c)} The means (points) and distributions (violin plots) of the circuit polarizations versus benchmark depth ($d$), and fits to an exponential $Ap^d$ (curves). \textbf{(d)} Error rates ($r$) obtained from the fit's decay rate for DRB and MRB versus the number of qubits ($n$), and the values predicted from calibration data. The DRB and MRB error rates are in close agreement, validating MRB against the reliable but unscalable DRB protocol. The measured $r$ diverges from the predictions of Quito's calibration data as $n$ increases, indicating crosstalk. \textbf{(e)} The mean polarizations at $d=0$ ($S_0$) decrease rapidly with $n$ for DRB and standard RB [at best $\log(S_0) = 1-O(n^2/\log n)$] making them infeasible beyond a few qubits, whereas $\log(S_0) = 1-O(n)$ for MRB. }
\label{fig:experiment}
\end{figure}

\begin{figure}[t!]
\includegraphics[width=8.5cm]{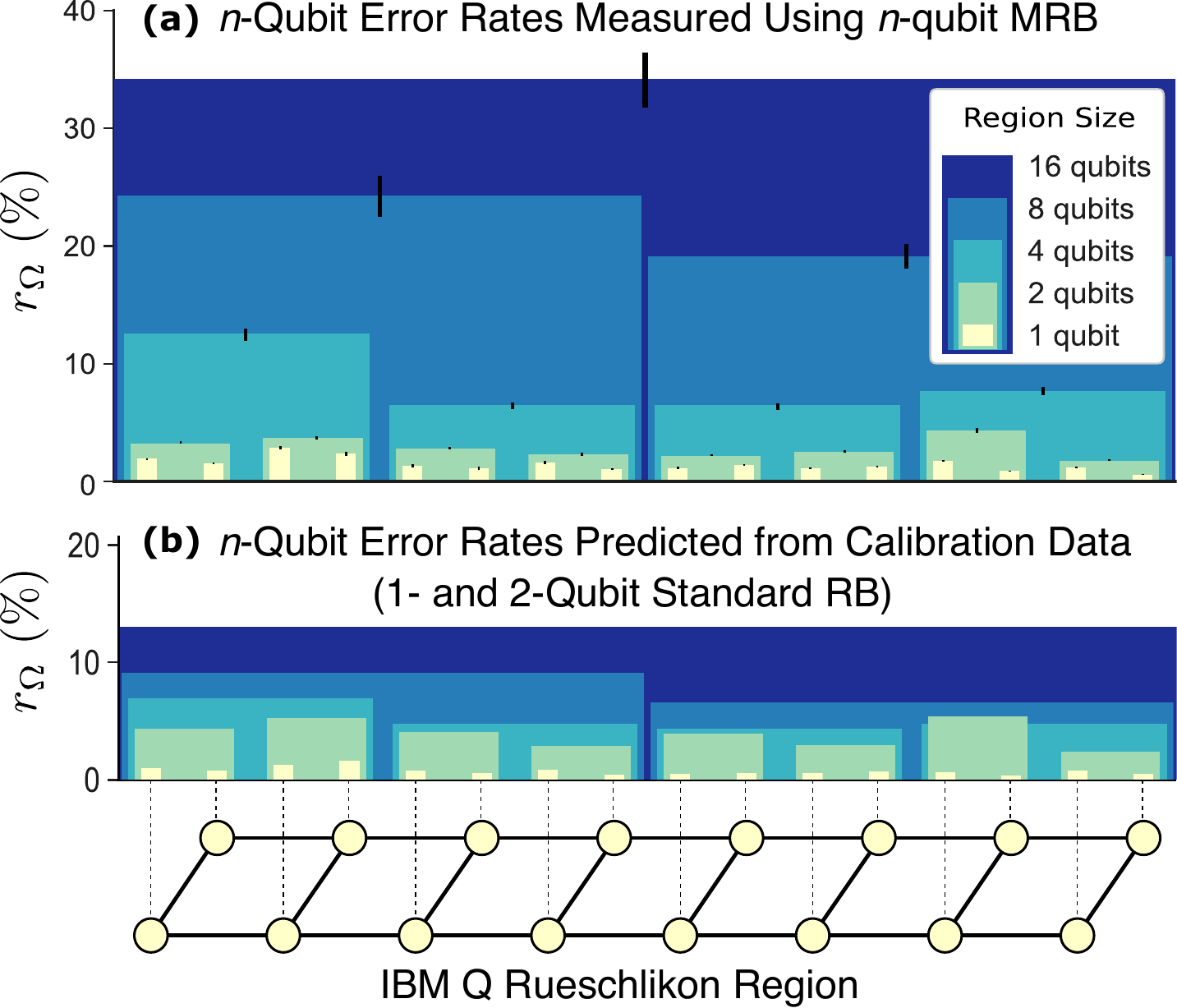}
\caption{\textbf{Mapping out the performance of a 16-qubit processor.} MRB with was used to probe the performance of $n$-qubit regions of IBM Q Rueschlikon. \textbf{(a)} The measured error rate $(r_{\Omega})$ for each qubit subset that was tested (black lines are $1\sigma$ uncertainties) and \textbf{(b)} the over-optimistic predictions from calibration data. The horizontal axis is a device schematic (nodes are qubits and edges the available CNOTs).}
\label{fig:experiment-2}
\end{figure}

In this Letter we introduce a simple, flexible, and robust RB method that removes the Clifford compilation bottleneck that limits current methods. We show how \emph{randomized mirror circuits} (Fig.~\ref{fig:1}a) enable scalable RB of Clifford gates. This work advances \emph{circuit mirroring} \cite{proctor2020measuring}, a recently introduced method for scalable benchmarking of quantum computers. Ref.~\cite{proctor2020measuring} shows how mirror circuits can be used to map out how a quantum computer's performance on circuits depends on their widths and depths (volumetric benchmarking \cite{blume2019volumetric}), but it doesn't show how to quantify gate fidelity. Here, we show how to use randomized mirror circuits to estimate the infidelity of an average Pauli-dressed \cite{Knill2005-xm, Wallman2016-rd, Ware2018-tc,  Hashim2020-qg} $n$-qubit circuit layer (Fig.~\ref{fig:1}a, grey boxes), and we present a theory that proves that this method---mirror RB (MRB)---is reliable. MRB can be applied whenever a typical $n$-qubit circuit layer has significantly non-zero fidelity, enabling RB of hundreds or even thousands of qubits  with physically realistic error rates [$O(10^{-2})$-$O(10^{-3})$]. We demonstrate and validate MRB on up to 225 qubits using simulations (Fig.~\ref{fig:sims}), and on up to 16 physical qubits using IBM Q's cloud quantum computing platform (Figs.~\ref{fig:1},~\ref{fig:experiment} and ~\ref{fig:experiment-2}).

\vspace{0.2cm}
\noindent
\textbf{Randomized mirror circuits.} MRB uses \emph{randomized mirror circuits} \cite{proctor2020measuring}, shown in Fig.~\ref{fig:1}a. By design, each randomized mirror circuit $C$ should ideally always produce a single bit string $s_C$ that is efficient to compute. Distributions over these circuits are parameterized by an $n$-qubit layer set $\mathbb{L}=\{L\}$ \footnote{$n$-qubit layers are also known as cycles or $n$-qubit gates.}, a probability distribution $\Omega$ over $\mathbb{L}$, and a benchmark depth $d$ that specifies the number of Pauli-dressed layers in the circuit. Both $\mathbb{L}$ and $\Omega$ are customizable, but we require that (1) each layer contains only Clifford gates, (2) each layer's inverse $L^{-1}$ is also within $\mathbb{L}$, (3) $\Omega(L)=\Omega(L^{-1})$, and (4) $\Omega$-random layers quickly locally randomize an error (local ``twirling'') and spread it across multiple qubits. Condition (4) is also required for reliable DRB, and the circumstances under which it is satisfied have been studied in detail \cite{proctor2018direct}. For all demonstrations herein, the layer set consists of parallel applications of CNOTs between connected qubits and all 24 single-qubit Clifford gates. This enables transparent quantification of the errors caused by native two-qubit gates, including crosstalk. Note, however, that our method can be applied to, e.g., CNOTs synthesized via SWAP chains, enabling comparisons between the errors in identical layer sets on different devices. All our distributions $\Omega$ have a similar form whereby sampling a layer consists of: (1) sampling some CNOTs, and (2) sampling uniformly random single-qubit Clifford gates for all qubits not acted on by those CNOTs.

\vspace{0.2cm}
\noindent
\textbf{Mirror RB.} MRB aims to measure $\epsilon_{\Omega} := \sum_{L} \Omega(L)\epsilon(L)$, where $\Omega$ is a user-chosen distribution over $\mathbb{L}$, and $\epsilon(L)$ is the entanglement infidelity of the Pauli-dressed version of the $n$-qubit layer $L$ (grey boxes, Fig.~\ref{fig:1}a). In all our demonstrations we do not compile the Paulis into the $L$ layers, but this is permissible. MRB estimates $\epsilon_{\Omega}$ using data from $\Omega$-sampled randomized mirror circuit. For each circuit $C$ that we run, we estimate its \emph{effective polarization}
\begin{equation}
S =  \frac{4^n}{4^n-1}\left[\sum_{k=0}^{n} \left(-\frac{1}{2}\right)^k h_k\right] - \frac{1}{4^n -1},
\label{eq:S}
\end{equation}
where $h_k$ is the probability that the circuit outputs a bit string that is a Hamming distance of $k$ from its target bit string ($s_C$).
As our theory (below) shows, the simple additional analysis in computing $S$ mitigates the limited ``twirling'' enacted by our circuits. 

MRB is the following protocol:
\begin{enumerate}
\item For a range of integers $d \geq 0$, sample $K$ randomized mirror circuits of benchmark depth $d$ where $d$ is even (see Fig.~\ref{fig:1}a), using the distribution $\Omega$, and run each one $N\geq 1$ times.

\item Estimate each circuit's effective polarization $S$.
\item Fit $\overline{S}_d$, the mean of $S$ at benchmark depth $d$, to $\overline{S}_d = Ap^d$, where $A$ and $p$ are fit parameters, and then compute $r_{\Omega} = (4^n - 1)(1 - p)/4^n$ as an estimate of $\epsilon_{\Omega}$.%
\end{enumerate}

\vspace{0.2cm}
\noindent
\textbf{Theory.} We now show that MRB is reliable, i.e., $\overline{S}_d \approx Ap^d$ and $r_{\Omega} \approx \epsilon_{\Omega}$ under broad conditions. We assume that errors are Markovian \cite{Nielsen2020-lt} but not necessarily gate-independent (many, but not all, non-Markovian errors appear Markovian within random circuits \cite{Hashim2022-pe, fong2017randomized, epstein2014investigating}). We use $U(L)$ and $\phi(L)$ to denote the $n$-qubit superoperators that represent a layer $L$'s perfect and imperfect implementations, respectively, and $\mathcal{E}(L)$ its error map, i.e., $\phi(L) =\mathcal{E}(L)U(L)$. Our theory starts from a single randomized mirror circuit $C$ of benchmark depth $d$. So $C=F_0^{-1}P_{d}L_1^{-1}\cdots P_{1+\nicefrac{d}{2}}L_{\nicefrac{d}{2}}^{-1} P_{\nicefrac{d}{2}}L_{\nicefrac{d}{2}}\cdots P_1L_1P_0F_0$, where (1) $P_i$ are Pauli layers, (2) $F_0$ and $F_0^{-1}$ consist of one-qubit Clifford gates, and (3) $L_i$ are $\Omega$-sampled layers and $L_i^{-1}$ their inverses. The components (1), (2) and (3) are sampled independently. To compute $\overline{S}_d$, as a function of $\mathcal{E}{(L)}$, we can therefore average over (1-3) separately in turn. 

The Pauli layers (green boxes, Fig.~\ref{fig:1}a) are independent, uniformly random, and interleaved between every other layer. They therefore have two effects: they randomize the target bit string ($s$), which guarantees that $\overline{S}_d \to 0$ as $d\to \infty$ to a good approximation \cite{harper2019statistical}, and they twirl the errors on the $L_i$ layers into stochastic Pauli errors \cite{Knill2005-xm, Wallman2016-rd, Ware2018-tc,  Hashim2020-qg}. So we can analyze the ``residual'' circuit $C=F_0^{-1}L_1^{-1}\cdots L_d^{-1}L_d \cdots L_1F_0$ with each $L_i$'s error map $\mathcal{E}(L_i)$ a stochastic Pauli channel. The composite superoperator for this circuit is $\phi(C) = \phi(F_0^{-1}) \mathcal{E}_d\phi(F_0)$ where $\mathcal{E}_d \equiv \phi(L_1^{-1})\cdots \phi(L_d^{-1}) \phi(L_d) \cdots  \phi(L_1) $ is a stochastic Pauli channel, as each $U(L_i)$ is a Clifford operator.

The intial layer (blue boxes, Fig.~\ref{fig:1}a) $F_0$ contains independent, uniformly random single-qubit Clifford gates. Averaging over this implements \emph{local} 2-design twirling on each qubit \cite{gambetta2012characterization}. That is, $\overline{\mathcal{E}}_d \equiv \frac{1}{24^n}\sum_{F_0}[U(F_0^{-1})\mathcal{E}_dU(F_0)]$ is a stochastic Pauli channel with equal \emph{marginal} probabilities to induce an $X$, $Y$ or $Z$ error on any fixed qubit. An error induced by $\overline{\mathcal{E}}_d$ flips at least one output bit iff it applies $X$ or $Y$ to at least one qubit. So, if $\overline{\mathcal{E}}_d$ induces a weight $k$ error (an error on $k$ qubits) the circuit outputs $s_C$ with a probability of $\nicefrac{1}{3}^k$. Generally, a weight $k$ error causes flips on $j$ of the output bits with probability $M_{jk}= {k \choose j} \frac{2^j}{3^k}$. So $\vec{h} = M \vec{p}$ where $h_k$ and $p_k$ are the probabilities that $k$ bits are flipped and that $\overline{\mathcal{E}}_d$ induces a weight $k$ error, respectively, with $k=0,\dots,n$. By inverting $M$, we obtain $p_0 = \sum_{k=0}^{n} \left(-\nicefrac{1}{2}\right)^kh_k \equiv H$. Because $p_0 = 1-\epsilon(\mathcal{E}_d)$ where $\epsilon(\mathcal{E}_d)$ is $\mathcal{E}_d$'s entanglement infidelity, $H$ therefore equals $\mathcal{E}_{d}$'s entanglement fidelity, and $S$ [Eq.~\eqref{eq:S}] its \emph{polarization} $\gamma(\mathcal{E}_d) := 1 - 4^n \epsilon(\mathcal{E}_d)/(4^n-1)$. State preparation and measurement (SPAM) errors also contribute to $S$ (and $H$), as do errors in $F_0$ and $F_0^{-1}$. But their effect is approximately $d$-independent, so $S\approx A \gamma(\mathcal{E}_d)$ for some $A$.

We have related a randomized mirror circuit's $S$ to the polarization of its superoperator [$\gamma(\mathcal{E}_d)$]. Now we relate $\gamma(\mathcal{E}_d)$ to the polarizations of the circuit's constituent layers [$\gamma(\mathcal{E}(L_i))$]. If every $\mathcal{E}(L_i)$ is an $n$-qubit depolarizing channel, with layer-dependent error rates, then $\gamma(\mathcal{E}_d) = \prod_{i=1}^{d} \gamma_{i^{-1}} \gamma_i$ where $\gamma_i \equiv \gamma(\mathcal{E}(L_i))$. More generally we argue that  $\gamma(\mathcal{E}_d) \approx \prod_{i=1}^{d} \gamma_{i^{-1}} \gamma_i$. For two stochastic Pauli channels $\mathcal{E}_A$ and $\mathcal{E}_B$, $\gamma(\mathcal{E}_A\mathcal{E}_B)=\gamma(\mathcal{E}_A)\gamma(\mathcal{E}_B) + \eta$ where $\eta = \sum_{j}(\epsilon_{A,j}-\frac{\epsilon[\mathcal{E}_A]}{4^n-1})(\epsilon_{B,j}-\frac{\epsilon[\mathcal{E}_B]}{4^n-1}])$ and $\vec{\epsilon}_{i}$ is the vector of $4^n-1$ Pauli error probabilities for $\mathcal{E}_i$. $\eta$ quantifies the rate that errors cancel when composing the two channels, relative to the rate that they cancel when composing $n$-qubit depolarization channels. It is negligible unless $\vec{\epsilon}_{A}$  and $\vec{\epsilon}_{B}$ are sparse (e.g., if $\vec{\epsilon}_{A}=\vec{\epsilon}_{B}$ and the error probability is equally distributed over $K$ errors, then $\eta = \epsilon(\mathcal{E}_A)^2\left[\frac{1}{K} - \frac{1}{4^n-1}\right]$). So, unless the Pauli error probability distributions of the $L_i$ are sharply spiked, then $\gamma(\mathcal{E}_d) \approx \prod_{i=1}^{d} \gamma_{i^{-1}} \gamma_i$ for any randomized mirror circuit. Furthermore, because of the properties that we demand of $\Omega$ (see above), our circuits are ``scrambling''---they locally randomize errors, and quickly spread them across many qubits. This suppresses error cancellation further \cite{proctor2018direct}. So  $\gamma(\mathcal{E}_d) \approx \prod_{i=1}^{d} \gamma_{i^{-1}} \gamma_i$ for a typical randomized mirror circuit.

Finally, we calculate the effect of averaging over the $L_i$ layers (pink boxes, Fig.~\ref{fig:1}a). They are independently sampled from $\Omega$, so $\overline{S}_d \approx A (\sum_L\Omega(L) \gamma_{L^{-1}} \gamma_L)^{\nicefrac{d}{2}}$ where $\gamma_L \equiv \gamma(\mathcal{E}(L))$. That is, $\overline{S}_d\approx Ap^d$ where $p^2 \approx \sum_L \Omega(L)\gamma_{L^{-1}}\gamma_L$. 
Rewriting this in terms of $\epsilon_{\Omega}$ and $\text{Cov}_{\Omega} =[\sum_{L}\Omega(L)\epsilon(L^{-1})\epsilon(L)]-[\epsilon_{\Omega}]^2$ gives $p^2 \approx (1-\frac{4^n}{4^n-1}\epsilon_{\Omega})^2 + \frac{4^n}{4^n-1} \text{Cov}_{\Omega}$. So if $\text{Cov}_{\Omega}=0$ then $r_{\Omega} \approx \epsilon_{\Omega}$. $\text{Cov}_{\Omega}$ quantifies the correlation between the error rate of a $\Omega$-random layer $L$ and its inverse $L^{-1}$, so $\text{Cov}_{\Omega} \neq 0$ is likely. This covariance satisfies $\epsilon_{\Omega}(1-\epsilon_{\Omega}) \geq \text{Cov}_{\Omega} \geq -\epsilon_{\Omega}^2$, so $ \epsilon_{\Omega} + O(\epsilon_{\Omega}^2) \gtrapprox r_{\Omega} \gtrapprox \frac{\epsilon_{\Omega}}{2} + O(\epsilon_{\Omega}^2)$. Therefore $r_{\Omega}$ is never significantly large than $\epsilon_{\Omega}$, and it can be smaller by at most a factor of $\approx 2$. The $\{\epsilon(L)\}$ distributions that get close to these bounds on $\text{Cov}_{\Omega}$ are not physically typical, e.g., the upper bound is saturated if $\epsilon(L)=\epsilon(L^{-1})$ and $\epsilon(L) =0$ or $\epsilon(L)=1$ for each $L$. We therefore conjecture that, for physically relevant $\{\epsilon(L)\}$, $r_{\Omega}$ typically only slightly underestimates $\epsilon_{\Omega}$. This is supported by our simulations and our demonstrations on physical qubits.

\vspace{0.2cm}
\noindent
\textbf{Simulations.} We simulated MRB on 1-225 qubits with randomly sampled stochastic Pauli error models. The qubits were arranged on a $15\times 15$ lattice (the layer set is described above). We independently sampled a total of 900 MRB circuit sets with a range of $n \in [1..225]$. We used a distribution $\Omega$ whereby a layer sampled from $\Omega$ has an expected CNOT density of $\nicefrac{1}{8}$. For each MRB circuit set we used a different randomly sampled error model, consisting of biased and correlated Pauli errors with one- and two-qubit gates having an expected infidelity of 0.1\% and 1\%, respectively \cite{supplement}. Fig.~\ref{fig:sims}a shows $\epsilon_{\Omega}$ versus $r_{\Omega}$. We observe that $r_{\Omega} \approx \epsilon_{\Omega}$, with $r_{\Omega}$ typically slightly less than $\epsilon_{\Omega}$, as expected from our theory. Quantifying estimation error by $\delta_{\rm rel} = \frac{r_{\Omega} -\epsilon_{\Omega}}{\epsilon_{\Omega}}$,  we find that $\delta_{\rm rel} > -0.32$ in all 900 simulations and for each $n$ its mean $\overline{\delta}_{\rm rel}$ satisfies $0.003>\overline{\delta}_{\rm rel}>-0.16$. Although this systematic underestimation of $\epsilon_{\Omega}$ is undesirable, it is arguably small enough to be insignificant (RB is typically used for rough estimates of gate performance rather than precision characterization).

To show how MRB can be used to reveal crosstalk errors, we simulated it on our hypothetical 225-qubit processor with two illustrative models, one with and one without crosstalk. The crosstalk-free model consisted of 0.5\% readout error on each qubit, and depolarization on the one- and two-qubit gates, with 0.1\% and 1\% error rates, respectively. In the crosstalk model, each CNOT also caused the error probability for qubit $q$ to increase by $\epsilon(q)$, with $\epsilon(q)$ a slowly decreasing function of the distance (on the lattice) from $q$ to the CNOT's location \cite{supplement}. Fig.~\ref{fig:sims}b shows $r_{\Omega}$ (points) and $\epsilon_{\Omega}$ (dotted line) versus $n$ for both models. We find that $r_{\Omega} \approx \epsilon_{\Omega}$ (averaged over $n$, $\overline{\delta}_{\rm rel} \approx -0.17$ and $\overline{\delta}_{\rm rel} \approx -0.08$ for the crosstalk-free and crosstalk models, respectively), and that $r_{\Omega}$ grows quadratically at low $n$ under the crosstalk model---an effect that cannot be observed without running many-qubit circuits.

\vspace{0.2cm}
\noindent
\textbf{Validating MRB with cloud access experiments.} To demonstrate MRB and compare it to existing techniques, we ran MRB, DRB \cite{proctor2018direct} and standard RB \cite{magesan2011scalable} on 1-5 qubits of IBM Q Quito \cite{ibmq}. DRB is designed to measure the same quantity as MRB ($\epsilon_{\Omega}$) and is known to be reliable but unscalable (because its circuits start by preparing a random $n$-qubit stabilizer state). For DRB and MRB we sampled layers with an expected CNOT density of $\xi = \nicefrac{1}{8}$ \cite{supplement} (standard RB does not have flexible sampling and its error rate is incomparable). Fig.~\ref{fig:experiment}a-c shows that we observe exponential decays for all three methods and all $n$ (for all methods $d=0$ corresponds to the shortest allowed circuit, consisting of a random $n$-qubit Clifford and its inverse for standard RB and preparation in and measurement of a random stabilizer state for DRB). For standard RB and DRB we rescale the success probabilities $P$ to polarizations $(P-\nicefrac{1}{2^n})/(1-\nicefrac{1}{2^n})$ [this has no effect on the estimated $r$] for easier comparison with MRB.

Fig.~\ref{fig:experiment}a-c also highlights the fundamentally improved scaling of  MRB. The $d=0$ polarization (Fig.~\ref{fig:experiment}e) decays much more quickly with $n$ for DRB and standard RB, because they use subroutines containing $O(\nicefrac{n^2}{\log n})$ gates, whereas $d=0$ randomized mirror circuits use $O(n)$ gates. The error rates estimated by DRB and MRB are in close agreement for all $n$ (Fig.~\ref{fig:experiment}d), validating MRB. We also predicted $r_{\Omega}$ from Quito's calibration data \cite{supplement}. These predictions (stars, Fig.~\ref{fig:experiment}d) are consistent with our observations for $n=1,2$, but they are over-optimistic as $n$ increases. This discrepancy indicates crosstalk errors caused by CNOTs. This is because IBM's one- and two-qubit calibration data are obtained from simultaneous one-qubit RB and isolated two-qubit RB (i.e., all other qubits are left idle) \cite{gambetta2012characterization, ibmq}, respectively. Therefore, the one-qubit error rates include contributions from any one-qubit gate crosstalk, whereas the two-qubit error rates do not include contributions from two-qubit gate crosstalk.

\vspace{0.2cm}
\noindent
\textbf{Mapping out a processor's performance.} MRB can be used to map out performance of a processor's $n$-qubit layers when varying both $n$ and the embedding of those qubits, as we demonstrate on IBM Q Rueschlikon (16 qubits) \cite{ibmq}. For $n \in \{1,2,4,8,16\}$ we divided Rueschlikon into $\nicefrac{16}{n}$ regions, and ran randomized mirror circuits on each region (the one-qubit circuits were performed simultaneously to match IBM's calibration experiments) \cite{supplement}. In this demonstration, we fixed the expected number of CNOTs in a layer to $\nicefrac{1}{2}$. Fig.~\ref{fig:1}b shows exponential decays for one region of each size (the leftmost regions in Fig.~\ref{fig:experiment-2}), and Figs.~\ref{fig:experiment-2}a and \ref{fig:experiment-2}b show $r_{\Omega}$ for all benchmarked regions and the predictions from the calibration data, respectively. The prediction underestimates $r_{\Omega}$ for $n > 2$, again signifying crosstalk induced by CNOTs (see discussion above).

\vspace{0.2cm}
\noindent
\textbf{Discussion.} In this Letter we have introduced a technique that enables holistic RB of hundreds or thousands of qubits, while retaining the core simplicity of standard RB---fitting data from random circuits to an exponential. We anticipate that techniques based on standard RB \cite{magesan2012efficient, harper2017estimating, chasseur2017hybrid, wallman2015estimating, feng2016estimating, sheldon2016characterizing, wood2017quantification,chasseur2015complete,wallman2015robust, rol2017restless, kelly2014optimal, gambetta2012characterization, McKay2020-no} can be enhanced using ideas introduced here. For example, MRB does not require compilation of subroutines so it removes the circuit scheduling complexities that plague simultaneous standard RB \cite{gambetta2012characterization, McKay2020-no}, suggesting that MRB will be more powerful for probing crosstalk. Similarly, running multiple MRB experiments with $\Omega$ varied could be used to isolate the error rates of different subsets of layers \cite{proctor2018direct}. This would enable reliable predictions of the performance of many-qubit, randomly-compiled circuits \cite{Knill2005-xm, Wallman2016-rd, Ware2018-tc,  Hashim2020-qg} (randomized compiling guarantees that layer fidelities are sufficient to predict overall circuit performance \cite{Hashim2022-pe}, which is not true otherwise \cite{proctor2020measuring}).

Our demonstrations on a cloud quantum computing platform revealed and quantified crosstalk errors that are invisible to one- and two-qubit RB, highlighting the need for scalable methods like ours. Outside the paradigm of RB there are a variety of methods for testing $n$-qubit circuit layers, and our technique complements them. For example, cycle benchmarking \cite{erhard2019characterizing, Hashim2020-qg} and Pauli noise estimation \cite{harper2019efficient, flammia2019efficient} can characterize a Pauli-dressed $n$-qubit layer. These techniques extract more information about a layer's errors, but, unlike MRB, they test only one (or a few) of a processor's many possible $n$-qubit layers. Methods for extracting more information from mirror circuit data, e.g., by using the techniques of Refs.~\cite{erhard2019characterizing, harper2019efficient, flammia2019efficient}, are an intriguing possibility \cite{Flammia2021-dn, Maksymov2021-ly}.

Our method is built on a particular type of randomized mirror circuits, but circuit mirroring \cite{proctor2020measuring} is a flexible tool that could be used to construct a range of randomized benchmarks with complementary properties to ours. For example, mirror circuits can contain non-Clifford gates \cite{proctor2020measuring}, which suggests a route to scalable RB of universal gate sets, and scalable ``full stack'' benchmarks. 

\vspace{0.2cm}
Since the completion of this manuscript, Mayer \emph{et al.}~\cite{Mayer2021-Theory} presented a complementary theory for MRB that assumes gate-independent errors and a 2-design gate set.

\begin{acknowledgments}
This work was supported by the Laboratory Directed Research and Development program at Sandia National Laboratories and the U.S. Department of Energy, Office of Science, Office of Advanced Scientific Computing Research through the Quantum Testbed Program. Sandia National Laboratories is a multi-program laboratory managed and operated by National Technology and Engineering Solutions of Sandia, LLC., a wholly owned subsidiary of Honeywell International, Inc., for the U.S. Department of Energy's National Nuclear Security Administration under contract DE-NA-0003525. All statements of fact, opinion or conclusions contained herein are those of the authors and should not be construed as representing the official views or policies of the U.S. Department of Energy, or the U.S. Government, or the views of IBM. We thank the IBM Q team for technical support.

\vspace{0.2cm}
All data and analysis code are available at 10.5281/zenodo.5197714. Our circuit sampling code is available in \texttt{pyGSTi} \cite{nielsen2020probing, pygstiversion0.9.10}.
\end{acknowledgments}
\bibliography{Bibliography}

\section*{SUPPLEMENTAL MATERIAL}
\subsection*{Simulations}
Here we provide details of the simulations of mirror RB presented in the main text (see Fig.~2). All the error models that we simulated consisted of only stochastic Pauli errors. This enabled simulations that are efficient (polynomial) in the number of qubits ($n$). We implemented ``weak'' simulation using a standard stochastic unravelling, i.e., for error model $M$ and circuit $C$, we sampled from $\mathsf{P}(x \mid C,M)$, denoting $C$'s output distribution over bit strings under error model $M$. Under our error models, for each layer in $C$, $M$ implies a probability distribution over the $n$-qubit Pauli operators (encoding a stochastic Pauli channel) that specifies the probability of each Pauli occurring after that layer (so the probability of the identity is the probability of no error). To obtain a sample from $\mathsf{P}(x \mid C,M)$, we sampled a Pauli operator instance after each layer in $C$ (from $M$'s error distributions) to obtain a sequence of Clifford operators. We then simulated that sampled sequence using \texttt{CHP} \cite{aaronson2004improved} (using an interface built into \texttt{pyGSTi} \cite{nielsen2020probing, pygstiversion0.9.10}), obtaining a single bit string. All of our simulations consisted of generating $N=100$ samples for each distinct circuit.

In the main text we presented the results of two distinct sets of simulations, shown in Figs.~2a and ~2b. In both cases, the simulations are of $n$-qubit mirror RB for 18 different values of $n$, ranging from $n=1$ up to $n=225$ qubits with exponential spacing. These simulations are of an $n$ qubit subset of $225$ qubits that are arranged on a $15 \times 15$ square lattice with CNOT gates available between nearest-neighbor qubits. We independently sampled 50 distinct mirror RB experiment designs (a set of mirror RB circuits) at each $n$, with $K=30$ circuits per benchmark depth. The randomized mirror circuits depend on two user-specified components: the $n$-qubit layer set $\mathbb{L}$ and the sampling distribution $\Omega$ over $\mathbb{L}$. The layer set that we used was constructed in the same way as in all of our demonstrations: $\mathbb{L}$ consisted of all parallel applications of CNOTs between connected qubits and all 24 single-qubit Clifford gates. The sampling distribution was the ``edge grab'' sampler, with a two-qubit gate density of $\nicefrac{1}{8}$, that was introduced and described in detail in the supplemental material of Ref.~\cite{proctor2020measuring}.

Fig.~2a shows the results of simulating each of these $50 \times 18 = 900$ mirror RB experiment designs once, with an independent, randomly sampled error model for each mirror RB experiment design. For each RB experiment design, an error model was sampled as follows. The readout for each qubit is assigned a bit flip error probability, sampled uniformly and independently for each qubit from the interval $[0\%,1\%]$. For each qubit, each of the 24 single-qubit gates (from the set of all 24 single-qubit Clifford gates) is independently assigned an $n$-qubit error map that acts after the gate. This error map has $3+3k$ error rate parameters that we sample, where $k$ is number of neighbours for this qubit ($k=4$ unless the qubit is on the edge of the lattice). The error map contains the 3 weight-1 Pauli errors on the target qubit, with error rates of $\{l_i\}_{i=1}^{3}$, and the $3k$ weight-1 Pauli errors that can occur on neighbouring qubits, with rates $\{c_i\}_{i=1}^{3k}$.
These rates are sampled as follows. We sample $\gamma$ uniformly from the interval $[0\%,0.2\%]$, $\kappa$ uniformly from the interval $[0.5,1]$, $l_i'$ uniformly from the interval $[0,1]$ for $i=1,2,3$, and $c_i'$ uniformly from the interval $[0,1]$ for $i=1,\dots,3k$. If the gate is a $\sigma_z$-basis rotation we set $c_i'=0$ for all $i$. We then set $l_i = \gamma \kappa l_i'/\sum_j l_j' $ and $c_i \gamma (1-\kappa) c_i'/\sum_j c_j' $. Therefore, the total rate of errors that this gate induces on its target qubit is $\epsilon_{\rm targ} = \kappa \gamma$, and the total rate of errors that it induces on its neighbours is $\epsilon_{\rm nn} = \gamma(1-\kappa)$ --- except for $\sigma_z$-basis rotation where $\epsilon_{\rm nn} =0$ (we make this choice as $\sigma_z$-basis rotation do not typically involve a physical pulse). The error map for each CNOT gate (corresponding to each edge on the lattice) is constructed in the same way, except that (1) $\gamma$ is sampled uniformly from the $[0\%,2\%]$, and (2) there are 15 possible Pauli errors on the target qubit, and so $\{l_i\}_{i=1}^{15}$. In each of the 900 independent simulations, we extract $\epsilon_{\Omega}$ from the error model and we estimate $r_{\Omega}$ from the simulated data. Fig.~2a shows a scatter plot of $\epsilon_{\Omega}$ versus $r_{\Omega}$.

The simulations in Fig.~2b consisted of sampling one mirror RB experiment design for each $n$ under two different error models --- with and without long-range crosstalk errors. In the first model (model 1) 
\begin{itemize}
\item The readout on each qubit is subject to a bit flip error with a probability of 0.5\%.
\item Each one-qubit gate (any one of the 24 single-qubit Clifford gates) is followed by a one-qubit depolarizing channel with an entanglement infidelity of 0.1\%.
\item Each CNOT gate is followed by a two-qubit depolarizing channel with an entanglement infidelity of 1\%.
\end{itemize}
The second model (model 2) is the same as model 1 except that the CNOT gates also cause long-range crosstalk errors. In particular, each CNOT is still followed by a two-qubit depolarizing channel with an entanglement infidelity of 1\% on its two target qubits, $q_1$ and $q_2$, but it also causes local one-qubit depolarization on all other qubits. It applies a one-qubit depolarizing channel on qubit $q$ with an entanglement infidelity of $\delta(q)$, where $\delta(q)$ is a decreasing function of the distance (on the lattice) from $q$ to the CNOT location. The specific function used was $\delta(q) =0.0035 \times 0.999^{d(q,q_1,q_2)} $ where $d(q,q_1,q_2)$ is the minimum of the distance between $q$ and $q_1$ and the distance between $q$ and $q_2$. For each of the two models, at each $n$ we extract $\epsilon_{\Omega}$ from the error model and we estimate $r_{\Omega}$ from the simulated data. Fig.~2b shows $\epsilon_{\Omega}$ and $r_{\Omega}$ versus $n$ for both error models.

\subsection*{IBM Q Quito demonstrations} Our demonstration on IBM Q Quito (\texttt{ibmq\_quito} version 1.1.3, a 5-qubit Falcon r4T processor) was performed in June 2021. It consisted of running mirror, direct and standard RB on 1-5 qubits. Randomized mirror circuits are constructed using two user-specified components: the $n$-qubit layer set $\mathbb{L}$ and the sampling distribution $\Omega$ over $\mathbb{L}$. The layer set that we used in this demonstration was constructed in the same way as in all of our demonstrations: $\mathbb{L}$ consisted of all parallel applications of CNOTs between connected qubits and all 24 single-qubit Clifford gates. The sampling distribution was the ``edge grab'' sampler, with a two-qubit gate density of $\nicefrac{1}{8}$, that was introduced and described in detail in the supplemental material of Ref.~\cite{proctor2020measuring}. For each $n$, we used exponentially spaced benchmark depths $d$ (see Fig.~3). For each $n$ and $d$, we sampled and ran $K=30$ randomized mirror circuits.

Direct RB \cite{proctor2018direct} contains the same flexible sampling as mirror RB: the user specifies $\mathbb{L}$ and $\Omega$. Direct RB is designed to measure the same average layer infidelity ($\epsilon_{\Omega}$), but because direct RB and mirror RB are different protocols there is no a priori guarantee that their empirical error rates $r_{\Omega}$ will be equal. One of the motivations for running direct RB alongside mirror RB is to show that we obtain approximately the same $r_{\Omega}$ in both techniques.
For direct RB, we therefore choose to use the same $\mathbb{L}$ and $\Omega$ as in our concurrent mirror RB demonstration. 

For our purposes, an $n$-qubit direct RB circuit of benchmark depth $d$ consists of (1) a circuit that generates a uniformly random $n$-qubit stabilizer state, (2) a sequence of $d$ Pauli-dressed layers, consisting of a uniformly random Pauli layer and then a layer $L$ sampled from $\Omega$, and (3) a circuit that returns the qubits to a uniformly random computational basis state. To implement direct RB, it is necessary to compile the sub-circuits of steps (1) and (3) into the available gates (and it is the large circuits generated by this compilation when $n\gg 1$ that makes direct RB unscalable). We do so using the open-source compilers in \texttt{pyGSTi}. To analyze the direct RB data, we fit the mean success probability versus depth to $Ap^d + \nicefrac{1}{2^n}$, and then $r_{\Omega}=(4^n -1)(1-p)/4^n$ as with mirror RB. We are able to fix the asymptote to $\nicefrac{1}{2^n}$ because the target bit string is randomized \cite{harper2019statistical}. In Fig.~3 we present polarization decays, instead of success probability decays, but note that this is only a rescaling.

Our standard RB demonstrations follow the standard procedure for RB over the Clifford group \cite{magesan2011scalable, magesan2012characterizing}, except that we randomized the ideal output bit string by only inverting each sequence of random Clifford gates up to a randomly sampled Pauli operator. Although this is not current standard practice, it has been advocated for elsewhere --- because it means that the success probability decay will asymptote to $\nicefrac{1}{2^n}$ \cite{harper2019statistical}. In particular, and as with direct RB, we fit the mean success probability versus depth to $Ap^d + \nicefrac{1}{2^n}$ (for standard RB, we use the number of Clifford subroutines minus two to define the benchmark depth $d$, as then the shortest circuits allowed by all three methods correspond to $d=0$). Standard RB does not contain any flexible sampling, but it is necessary to compile the $n$-qubit Clifford subroutines in the circuits into the available gates. As with direct RB, we do so using the open-source compilers in \texttt{pyGSTi}. These compilations are unlikely to be optimal (e.g., in the number of CNOTs in the circuits), and so a slower decrease in the $d=0$ success probability with $n$ (or, equivalently, a slower decrease in the $d=0$ mean polarization $\overline{S}_0$, which is what is shown in Fig.~3e) than observed in our demonstration might have been achievable. However, at least $O(\nicefrac{n^2}{\log(n)})$ two-qubit gates are required to implement a typical $n$-qubit Clifford operator \cite{aaronson2004improved, patel2003efficient, bravyi2020hadamard}, so it is not possible to improve upon the scaling seen in our results. 

In the main text we do not present the error rates $r$ obtained from standard RB, as they are not directly comparable to the direct and mirror RB error rates. Specifically, standard RB measures (up to some subtleties \cite{proctor2017randomized, wallman2017randomized, merkel2018randomized, carignan2018randomized}) the infidelity of an average $n$-qubit Clifford gate, whereas direct and mirror RB approximately measure $\epsilon_{\Omega}$ --- the error rate of an average layer. The observed $n$-qubit Clifford RB error rate $r_n$ is $r_1=0.041(1)\%$, $r_2=1.97(5)\%$,  $r_3=17.9(7)\%$,  $r_4=67(6)\%$ and $r_5=95(7)\%$, where we have defined $r = (4^n-1)(1-p)/4^n$ rather than using the more common definition of $r = (2^n-1)(1-p)/2^n$ for consistency with the convention that we have used for mirror and direct RB (which we choose due to the arguably preferable properties of entanglement infidelity compared to average gate infidelity).

\subsection*{IBM Q Rueschlikon demonstrations}
Our demonstration on IBM Q Rueschlikon (\texttt{ibmqx5} version 1.1.0, a 16-qubit processor) was performed in July 2018. It consisted of running mirror RB on multiple $n$-qubit subsets of the processor, for $n \in \{1,2,4,8,16\}$. For each $n$, we divided Rueschlikon into $\nicefrac{16}{n}$ regions (as shown in Fig.~4), and we ran randomized mirror circuits on each region. The one-qubit circuits were performed simultaneously to match IBM's calibration experiments. These circuits were constructed using a sampling distribution $\Omega$ for which sampling a layer consisted of (1) selecting a pair of connected qubits uniformly at random, and adding a CNOT between those qubits to the layer with probability 50\%, and (2) sampling independent and uniformly random single-qubit Clifford gates to apply to all qubits that do not have a CNOT acting on them. A layer sampled from this $\Omega$ therefore contains no CNOT gates with 50\% probability and otherwise it contains one CNOT gate. Note that this sampling distribution is not suitable for much larger processors (unlike the edge grab sampler used in our simulations and the demonstrations on IBM Q Quito). This is because the CNOT density $\xi \to 0$ as $n\to\infty$ and circuits containing very few two-qubit gates cannot be scrambling \cite{proctor2018direct}. The data from these demonstrations was also presented in Ref.~\cite{proctor2020measuring} (see Fig.~1 therein), where it was used to construct volumetric benchmarking plots \cite{blume2019volumetric} rather than estimate layer error rates.
\end{document}